\documentclass[reprint,prd,twocolumn,superscriptaddress,showpacs,nofootinbib,floatfix,preprintnumbers]{revtex4-1}
\usepackage{graphicx, bm}
\newcommand{\lsim}{\mbox{\raisebox{-.6ex}{~$\stackrel{<}{\sim}$~}}}

\begin{document}
\preprint{MAN/HEP/2013/16}
\title{Probing TeV Left-Right Seesaw at Energy and Intensity Frontiers: \\ a Snowmass White Paper}

\author{P. S. Bhupal Dev}
\affiliation{Consortium for Fundamental Physics, School of Physics and Astronomy, University of Manchester, Manchester M13 9PL, United Kingdom}

\author{R. N. Mohapatra} 
\affiliation{Maryland Center for Fundamental Physics and Department of Physics, University of Maryland, College Park, Maryland 20742, USA}

\begin{abstract}
We discuss ways to probe the origin of neutrino masses at the Energy and Intensity frontiers, in TeV-scale Left-Right seesaw models where small neutrino masses arise via type-I seesaw mechanism. We consider generic  (`vanilla') version of such models as well as a version  which leads to `large' light-heavy neutrino mixing while keeping the neutrino masses small in a natural manner.  We point out specific observable implications of these classes of models at the LHC as well as in searches for lepton flavor violating processes such as $\mu\to e\gamma$ and  $\mu\to 3 e$, and also in searches for lepton number violating    
neutrinoless double beta decay. 
\end{abstract}
\maketitle
\section{Introduction}
The neutrino oscillation data unambiguously establish that neutrinos have tiny but non-zero masses as well as mixing between different flavors. Their understanding requires physics beyond the Standard Model (SM) since the SM predicts vanishing masses for neutrinos and hence no mixing. As the origin of mass for all charged fermions in the SM has now been clarified by the discovery of the Higgs boson, an important question is where does the neutrino mass come from. If we simply add three right-handed (RH) neutrinos ($\nu_R$) to the SM, one can write a Yukawa coupling of the form 
\begin{eqnarray}
{\cal L}_{\nu,Y}=h_{\nu, ij} \bar{L}_i \phi \nu_{R,j},
\end{eqnarray}
where $L_i=(\nu_i, \ell_i)_L^{\sf T}$ (with $i=e,\mu,\tau$) is the $SU(2)_L$ lepton doublet, and $\phi=(\phi^+,\phi^0)^{\sf T}$ is the SM Higgs doublet. After electroweak symmetry breaking via the Higgs vacuum expectation value (vev) $\langle \phi \rangle = (0,v)^{\sf T}$, this gives a Dirac mass to the light neutrinos: $m_\nu = h_\nu v$. To get sub-eV neutrino masses, it requires $h_\nu \simeq 10^{-12}$ which is an unnaturally small number. So the strong suspicion among theorists is that there is some new mechanism beyond the SM  which gives mass to the neutrinos in conjunction with the standard Higgs mechanism. This will also be a clue to the nature of physics beyond the SM.

A simple paradigm for small neutrino masses is the (type-I) seesaw mechanism~\cite{type1} where the RH neutrinos alluded to above have a Majorana mass, in addition to having a Dirac mass like all charged SM fermions. Neutrinos being electrically neutral allow for this novel possibility making them different from the charged fermions and this might be at the root of such diverse mass and mixing patterns.  This leads to the seesaw matrix with the following generic form in the $ (\nu_L, \nu_R)$ space:
\begin{eqnarray}
\left(\begin{array}{cc}0 & m_D\\ m^{\sf T}_D & M_{\nu_R}\end{array}\right)
\label{seesaw} \end{eqnarray}
where $m_D$ mixes the $\nu_L$ and $\nu_R$ states and is generated by the SM Higgs mechanism, whereas $M_{\nu_R}$ is the Majorana mass for $\nu_R$ which embodies the new neutrino mass physics.   In the usual  seesaw approximation $\|m_D\| \ll \|M_{\nu_R}\|$, where $\|M\|\equiv  \sqrt{{\rm Tr} (M^\dagger M)}$, the mass matrix for the light neutrinos is given by~\cite{type1}
\begin{eqnarray}
M_\nu \simeq -m_DM_{\nu_R}^{-1}m_D^{\sf T}, 
\label{mnu}
\end{eqnarray}
and a mixing between the heavy  and light neutrinos of order $V_{\ell \nu_R}\equiv m_DM^{-1}_{\nu_R}$~\cite{valle}. 

From (\ref{seesaw}), it is clear that there are two key aspects to the seesaw mechanism, i.e., the Majorana character of neutrino masses and the heavy-light neutrino mixing, that can be tested experimentally. 
A classic way to test the Majorana nature of both the light and heavy neutrinos is via neutrinoless double beta decay ($0\nu\beta\beta$) searches~\cite{rode}. However, this may not necessarily probe the heavy-light 
mixing since these mixing effects on $0\nu\beta\beta$ amplitude could be sub-dominant compared to those from purely LH or RH currents. The presence of heavy-light mixing can be explicitly searched for via departures from unitarity of the light neutrino mixing matrix~\cite{unitarity} in neutrino oscillation data as well as other low energy experiments (e.g., charged lepton flavor violation and leptonic $C\!P$ violation searches) at the Intensity frontier~\cite{Hewett}. But these observables do not prove the Majorana nature of heavy neutrinos since models with pseudo-Dirac neutrinos could also give rise to large 
non-unitarity effects~\cite{invunitarity}. On the other hand, both the Majorana nature and heavy-light mixing could manifest simultaneously at the Energy frontier via their distinct collider signals~\cite{atre}, thus giving complementary information to what is obtained from the low energy searches. Our goal in this paper is to explore this synergy between the Energy and Intensity frontiers which might provide a decisive test of the seesaw mechanism as the origin of neutrino masses and mixing.  
 A necessary requirement for this exploration to have any chance of success is that the seesaw scale be at most in the TeV range and the heavy-light mixing be relatively large. With this in mind, we discuss a class of models where these two requirements are satisfied in a somewhat natural manner. 

A natural class of models that provide an ultra-violet completion of seesaw is the Left-Right (L-R) symmetric theory of weak interactions based on the gauge group $SU(2)_L\times SU(2)_R\times U(1)_{B-L}$~\cite{LR}, where the main ingredients of seesaw, i.e. the RH neutrinos and their Majorana mass, appear naturally as follows: the RH neutrinos along with the RH charged leptons are assigned in a parity-symmetric way to the $SU(2)_R$ doublets $R_i=(\nu_i,  \ell_i)_R^{\sf T}$ -- the RH counterparts of the SM $SU(2)_L$ doublets $L_i$. The RH neutrinos are therefore a necessary part of the model and do not have to be added adhocly just to implement the seesaw mechanism. They acquire a Majorana mass as soon as the $SU(2)_R\times U(1)_{B-L}$ symmetry is spontaneously broken at a scale $v_R$. This is quite analogous to the way the charged fermions get mass as soon as the SM gauge symmetry $SU(2)_L\times U(1)_Y$ is broken at a scale $v$. The Higgs field that gives mass to the RH neutrinos becomes the RH analog of the SM Higgs boson. Thus the seesaw scale (synonymous with the RH neutrino mass) becomes intimately connected to the $SU(2)_R\times U(1)_{B-L}$ breaking scale. 

 
 
 Since L-R symmetric theories lead to new effects or add new contributions to already known low energy weak processes, it is necessary to know whether TeV-scale $SU(2)_R$-breaking is compatible with observations of low energy processes. It turns out that hadronic flavor changing neutral current effects such as $\epsilon_{K}$ and $K_L-K_S,~B_S-\overline{B}_S$ mixing receive significant contributions from RH charged current effects and therefore provide the most stringent constraints on $v_R$ by restricting the mass of the RH charged gauge boson $W_R$  to be $M_{W_R} \geq 2.5$ TeV~\cite{KL-KS}. Since at the Energy frontier the LHC energy goes up to 14 TeV, the L-R  model as a theory of neutrino mass can be probed as long as $M_{W_R}$ is below 5 - 6 TeV~\cite{goran}. There are also low energy tests of the model in the domain of leptonic physics such as the lepton flavor violating (LFV) $\mu\to e\gamma$ and $\mu\to 3e$ processes which are the focus of the Intensity frontier. Thus TeV scale L-R models provide a synergy between the Energy and Intensity frontiers.
 
 This article is organized as follows: in Section~\ref{sec2}, we give a brief overview of the generic TeV scale L-R models (model A) for seesaw mechanism. In Section~\ref{sec3} we describe a scenario (model B) where the Dirac mass matrix that goes into the seesaw formula has a specific texture that guarantees that neutrino masses are naturally small and the heavy-light mixing $V_{\ell N}$ is `large' enough to give new effects at the LHC and in LFV processes. In Section~\ref{sec4}, we give the implications of both models A and B for LHC searches via $W_R$ effects; in Section~\ref{sec5}, we present their implications for LFV, and  in Section~\ref{sec6} for $0\nu\beta\beta$. Our conclusions are given in Section~\ref{conc}.
\section{Minimal TeV Left-Right seesaw model}\label{sec2}
In the minimal L-R model, the fermions are assigned to the gauge group $SU(2)_L\times SU(2)_R\times U(1)_{B-L}$ as follows: denoting $Q\equiv (u,d)^{\sf T}$ and $\psi\equiv (\nu_\ell, \ell)^{\sf T}$ 
as the quark and lepton doublets respectively, we assign $Q_L$ and $\psi_L$ (also denoted by $L$) as the doublets under the $SU(2)_L$ group, while $Q_R$ and $\psi_R$ (also denoted by $R$) are the doublets under the $SU(2)_R$ group. The Higgs sector of the model consists of one or several of the following multiplets:
\begin{eqnarray}
\Delta_R\equiv\left(\begin{array}{cc}\Delta^+_R/\sqrt{2} & \Delta^{++}_R\\\Delta^0_R & -\Delta^+_R/\sqrt{2}\end{array}\right), ~\phi\equiv\left(\begin{array}{cc}\phi^0_1 & \phi^+_2\\\phi^-_1 & \phi^0_2\end{array}\right)
\end{eqnarray}
The gauge symmetry $SU(2)_R\times U(1)_{B-L}$ is broken by the vev $\langle \Delta^0_R\rangle = v_R$ to the group $U(1)_Y$ of the SM. 
There is also an LH counterpart ($\Delta_L$) to $\Delta_R$ which we do not consider here. There are versions of the model where parity and $SU(2)_R$ gauge symmetry scales are decoupled so that the $\Delta_L$ fields are absent from the low energy domain~\cite{CMP}. We will focus on this class of models here.
The vev of the $\phi$ field given by $\langle\phi\rangle={\rm diag}(\kappa, \kappa')$ breaks the SM gauge group to $U(1)_{\rm em}$.

To see how the fermions pick up mass and how seesaw mechanism arises, we write down the Yukawa Lagrangian of the model:
\begin{eqnarray}
{\cal L}_Y&=&h^{q,a}_{ij}\bar{Q}_{L,i}\phi_aQ_{R,j}+\tilde{h}^{q,a}_{ij}\bar{Q}_{L,i}\tilde{\phi}_aQ_{R,j}+
h^{\ell,a}_{ij}\bar{L}_i\phi_aR_j \nonumber \\
&&+ \tilde{h}^{\ell,a}_{ij}\bar{L}_i\tilde{\phi}_aR_j
+f_{ij} (R_iR_j\Delta_R +L_iL_j\Delta_L)+{\rm h.c.}
\end{eqnarray}
where $i,j$ stand for generations and $a$ for labeling Higgs bi-doublets, and $\tilde{\phi}=\tau_2\phi^*\tau_2$ ($\tau_2$ being the second Pauli matrix). After symmetry breaking, the Dirac fermion masses are given by the generic formula $M_f~=~h^f\kappa + \tilde{h}^f\kappa'$ for up-type fermions; for down quarks and charged leptons, it is the same formula with $\kappa$ and $\kappa'$ interchanged.  Thus we get the Dirac mass matrix  for neutrinos $m_D = h^{\ell}\kappa + \tilde{h^{\ell}}\kappa'$ and the Majorana mass matrix for the heavy RH neutrinos $M_{\nu_R}=fv_R$ which go into (\ref{seesaw}) for calculating the neutrino masses and heavy-light neutrino mixing.

In generic TeV scale seesaw models, i.e. without any special structure for $m_D$, we must fine-tune the magnitude of $m_D$ to be very small: $m_D\sim$ MeV for $M_{\nu_R}\sim$ TeV in order to get small neutrino masses. As a result, the seesaw structure~(\ref{seesaw}) implies that the heavy-light mixing parameter $V_{\ell \nu_R}\simeq \sqrt{M_\nu(fv_R)^{-1}}\lsim 10^{-6}$. This suppresses all heavy-light mixing effects to an unobservable level. We will call this the generic (`vanilla') L-R seesaw model (or model A).

\section{Models with 
enhanced heavy-light neutrino mixing: model B}\label{sec3}
As discussed above, for generic forms of $m_D$ and $M_{\nu_R}$, the heavy-light mixing parameter $V_{\ell \nu_R} \simeq \sqrt{M_\nu(fv_R)^{-1}}$ is a tiny number regardless of whether the seesaw scale is in the TeV range or higher. This keeps its effect shielded from being probed at either Energy or Intensity frontier. However, there are some special textures for $m_D$ for which even with TeV-scale seesaw, the mixing parameter $V_{\ell \nu_R}$ can be significantly enhanced whereas the neutrino masses remain naturally small. We present only one example here to illustrate our case, although several others have been discussed in the literature~\cite{cancel}. Consider the matrices $m_D$ and $M_{\nu_R}$ of the following form:
\begin{eqnarray}
m_D=\left(\begin{array}{ccc} a & \delta_1 & \epsilon_1\\ b & \delta_2 & \epsilon_2\\ c &\delta_3 & \epsilon_3\end{array}\right)~{\rm and}~
M_{\nu_R}=\left(\begin{array}{ccc} 0 & M_1 & 0\\M_1 & 0 & 0 \\ 0&0&M_2\end{array}\right)
\label{eq:texture}
\end{eqnarray}
with $\epsilon_i, \delta_i \ll a,b,c $. 
In the limit of $\epsilon_i, \delta_i\to 0$, the neutrino masses vanish, although the heavy-light mixing given by $V_{\ell \nu_{R,i}}=m/M_i$ (with $m=a,b,c$) can be quite large. The neutrino masses given by the seesaw formula (\ref{mnu}) become proportional to the products of  $\epsilon_i$ and $\delta_i$. If by some symmetry one can guarantee the smallness of $\delta_i$ and $\epsilon_i$, then  we have a TeV scale seesaw model with enhanced $V_{\ell \nu_R}$. These mass textures can be embedded into L-R models and have been shown to reproduce observed neutrino masses and mixing~\cite{dlm}. This is a highly non-trivial result since in L-R models the charged lepton mass matrix and the Dirac neutrino mass matrix are related, especially when there are additional discrete symmetries to guarantee the form of the Dirac mass $m_D$ given in Eq.~(\ref{eq:texture}). 
These models have important phenomenological implications  for collider signals, LFV effects, non-unitarity of the PMNS mixing matrix, etc. These considerations can reveal underlying symmetries of the lepton sector, which will be an important step towards a full understanding of the seesaw mechanism. We will call this the model B, and discuss in the following sections the implications of these scenarios for colliders and other low energy processes.  Note that while we have presented only one example of such non-generic Dirac mass matrix in Eq.~(\ref{eq:texture}),  our following results are derived in a model-independent phenomenological approach without restricting to a particular texture, and are also applicable to other Dirac textures discussed in literature~\cite{cancel}. Henceforth, we will generically take the heavy neutrino masses $M_{\nu_{R,i}}$ to be $M_N$  and the mixing $V_{\ell \nu_{R,i}}$ to be $V_{\ell N}$, and treat these two as free parameters. 

\section{Collider signatures of TeV scale Left-Right seesaw}\label{sec4}
\begin{figure*}[htb]
\centering
\begin{tabular}{cccc}
\includegraphics[width=4cm]{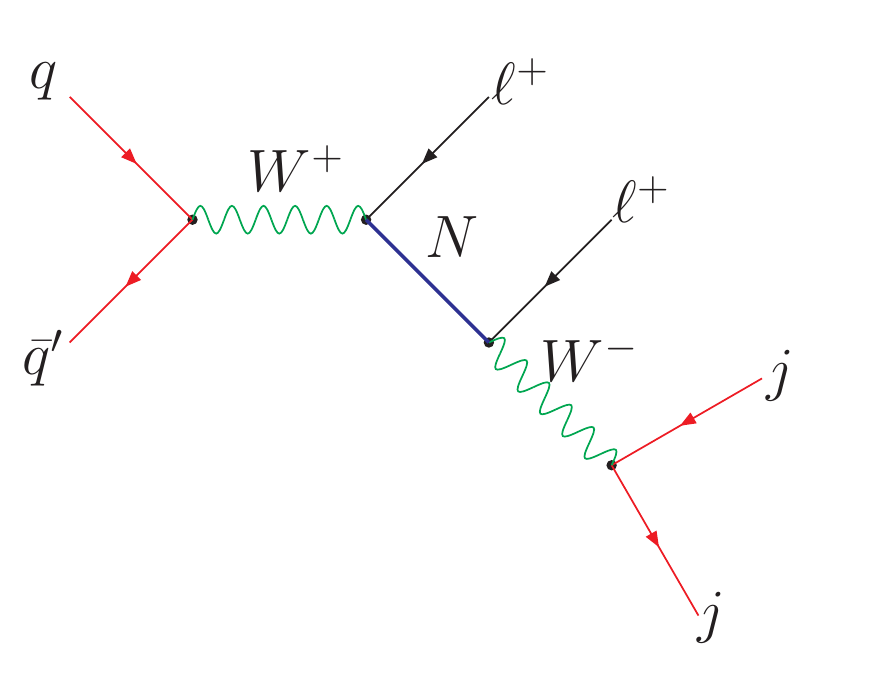} & 
\includegraphics[width=4cm]{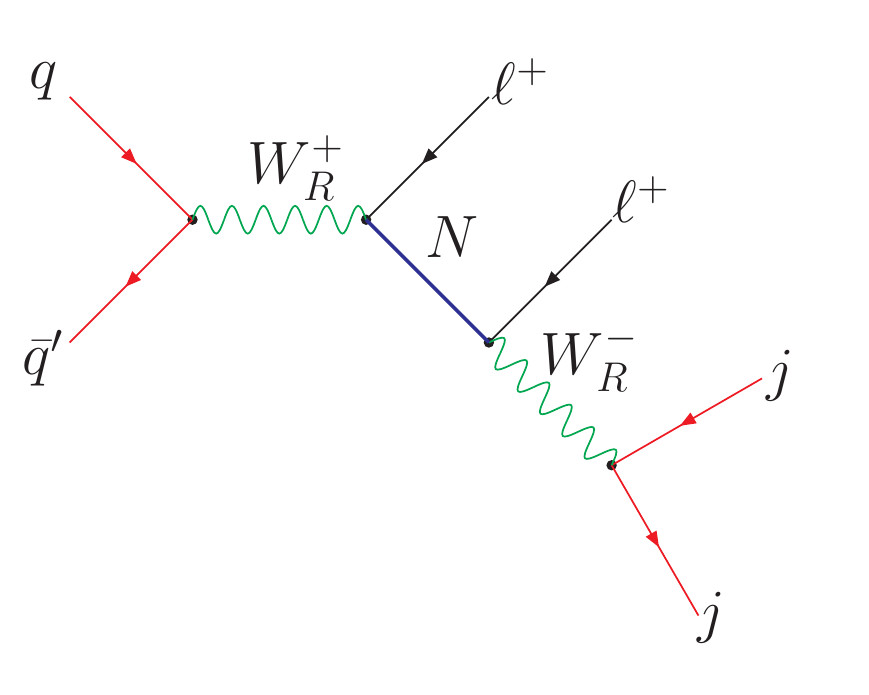} &
\includegraphics[width=4cm]{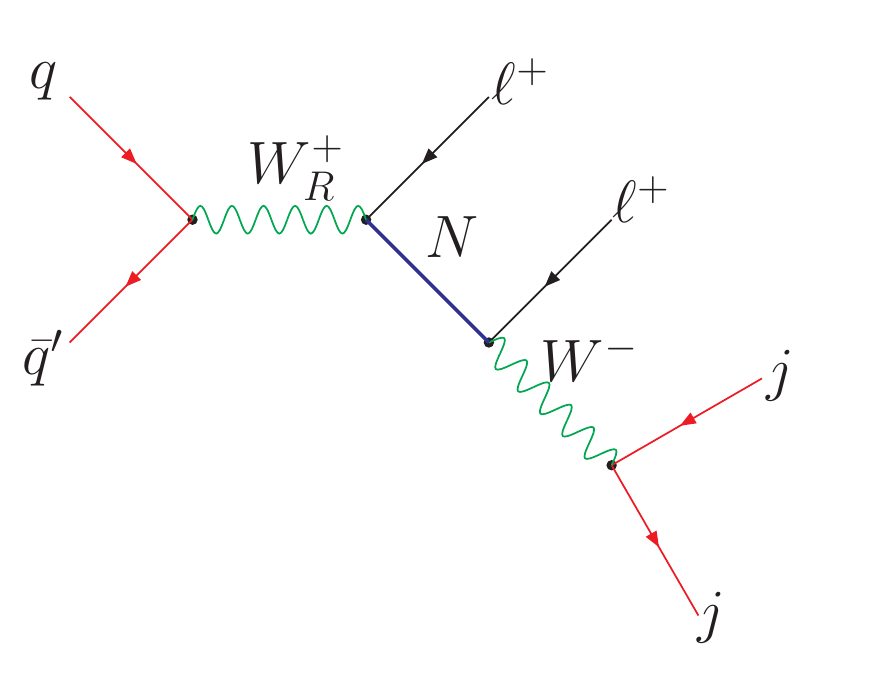} &
\includegraphics[width=4cm]{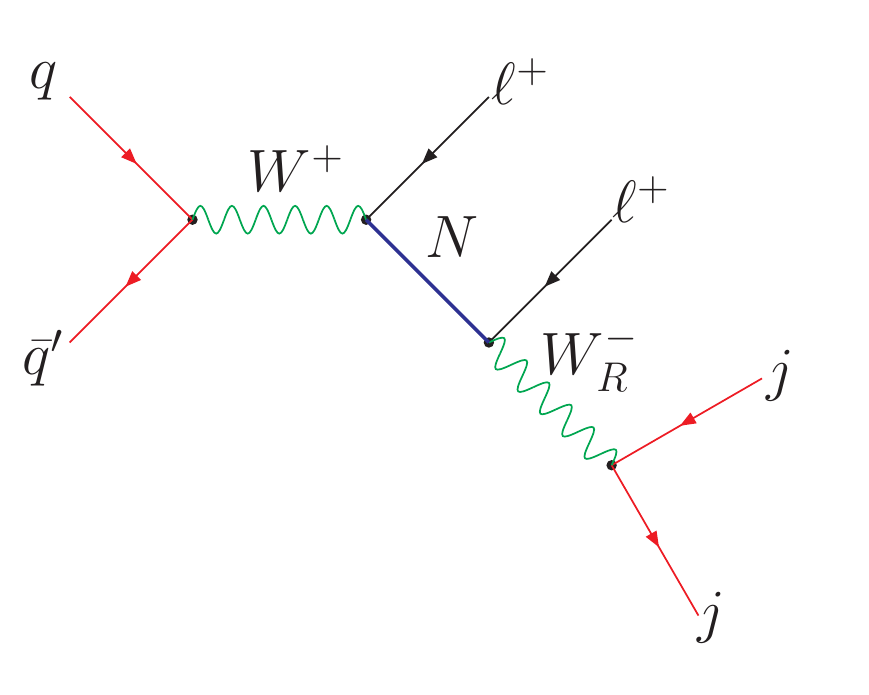} \\
(a) LL & (b) RR & (c) RL & (d) LR 
\end{tabular}
\caption{The Feynman diagrams contributing to the `smoking gun' collider signal $\ell^\pm\ell^\pm jj$ of a heavy Majorana neutrino in the minimal L-R seesaw model. }
\label{fig1}
\end{figure*}
There exist limits on the seesaw parameters $M_N$ and $V_{\ell N}$ from various low energy experiments as well as from LEP for $M_N<M_Z$~\cite{atre}. With the recent LHC Higgs data, these limits have been extended to $M_N\leq 200$ GeV~\cite{dfm}.
As far as the direct collider test of seesaw is concerned, we need a suitable combination of  the Majorana mass $M_{N}$ and mixing $V_{\ell N}$ which gives the `smoking gun' signal of same-sign dilepton plus two jets without missing energy ($\ell^\pm\ell^\pm jj$) at the LHC~\cite{KS}.  
In the absence of L-R symmetry, this signal depends crucially on the heavy-light neutrino mixing  (see Fig.~\ref{fig1}a) and can effectively probe the heavy neutrino masses $M_{N}$ only up to 300 GeV or so for $V_{\ell N}\geq 0.01$~\cite{theory-LL}. 
It must be stressed that any positive signal would not only signify the Majorana character of the heavy  neutrino  $\nu_R$ but also the non-generic structure of $m_D$ for reasons discussed in Sections \ref{sec2} and \ref{sec3}. 


 In the L-R symmetric embedding of TeV scale seesaw, the presence of RH gauge interactions could lead to significant enhancement for $\ell^\pm\ell^\pm jj$ signal, from $W_R$-mediated production and decay of $N$ (Fig.~\ref{fig1}b) as was first pointed in~\cite{KS} and from a combination of $V_{\ell N}$ and $W_R$ (Fig.~\ref{fig1}c) as recently pointed out~\cite{chendev}. In addition,  there are new contributions to $0\nu\beta\beta$ from $W_R$ exchange~\cite{MS} as well as new LFV effects~\cite{riaz, rnm-92, pal}. In this section, we explore the collider prospects, and in subsequent sections, we will discuss other low energy prospects of TeV L-R seesaw. 
 
 Important for the collider discussion is the relative value of $M_{W_R}$ and $M_N$. There are theoretical arguments based on vacuum stability which suggest that the heavy neutrinos in the minimal L-R seesaw models are lighter than the RH gauge bosons~\cite{rnm} for a large range of model parameters. We will therefore consider this mass ordering in this paper, although going beyond the minimal version, one could avoid this restriction. A major implication of this is that for RH gauge boson masses below 5 - 6 TeV, when it can be produced at the $\sqrt s=14$ TeV LHC with an observable cross section, its direct decay to on-shell RH neutrinos, which subsequently decay to the SM $W_L$-boson and charged leptons, will allow a probe of the heavy-light neutrino mixing for a wider mass range of up to a few TeVs from a study of  dilepton plus two jet final states~\cite{chendev}.  This recent result is discussed in somewhat details in the following two subsections.
\subsection{The Left-Right Phase Diagram}
There are four classes of Feynman diagrams in L-R symmetric models which can lead to the $\ell^\pm\ell^\pm jj$ final states (Fig. 1). We denote them as  (a) RR, (b) LL, (c) RL, and (d) LR, according to the chirality of the final state lepton-pair. The most widely studied of these are the LL and RR diagram in Fig.~\ref{fig1}a and Fig.~\ref{fig1}b respectively, the first in the context of simple seesaw~\cite{theory-LL} and the second in L-R models~\cite{KS,theory-RR}.
The channel in Fig.~\ref{fig1}a is a clear probe of the seesaw matrix even in the absence of L-R symmetry, although its effectiveness solely relies on the largeness of the heavy-light neutrino mixing. This channel was used to probe the mass range $M_N=100$ - 300 GeV at the LHC~\cite{CMS-LL, ATLAS-LL}, and direct upper limits on $|V_{\ell N}|^2$ of order 0.01 - 0.1 have been set from the $\sqrt s=7$ TeV LHC data. These limits could however be significantly improved (by about 50\%) if we include the infrared-enhanced production processes for the heavy neutrinos~\cite{dpy}. 
In the case of  Fig.~\ref{fig1}b, the heavy neutrinos are produced on-shell via the decay of $W_R$ and subsequently decay into a three-body final state via an off-shell $W_R$. Using this channel, LHC exclusion limits were derived in the $M_N$ - $M_{W_R}$ plane~\cite{ATLAS-RR, CMS-RR}, excluding $M_{W_R}$ up to 2.5 TeV for a TeV-scale $M_N$. Note that these limits are independent of the Dirac neutrino Yukawa coupling characterizing the mixing between the LH and RH neutrinos and therefore do not probe the seesaw matrix. 

The contributions in Fig.~\ref{fig1}c and Fig.~\ref{fig1}d, on the other hand, necessarily involve the heavy-light neutrino mixing and are clear probes of seesaw matrix. As has been shown recently~\cite{chendev}, for a sizable fraction of the  allowed parameter space,  the RL channel shown in Fig.~\ref{fig1}c gives the dominant contribution to the heavy neutrino signal.\footnote{For the importance of this channel in other contexts, see e.g.,~\cite{Chen:2011hc} for  distinguishing a heavy Majorana neutrino at the LHC from a pseudo-Dirac one, and~\cite{Han:2012vk} for determining the chirality of a heavy gauge boson.} Using this information, in combination with the existing experimental bounds, improved constraints on the mixing parameter have been derived~\cite{chendev} which can become substantially better than those derived from the LL channel alone as we increase the heavy neutrino mass and/or center of energy at the LHC. 
\begin{figure}[t]
\vspace{0.2cm}
\centering
\includegraphics[width=4.1cm]{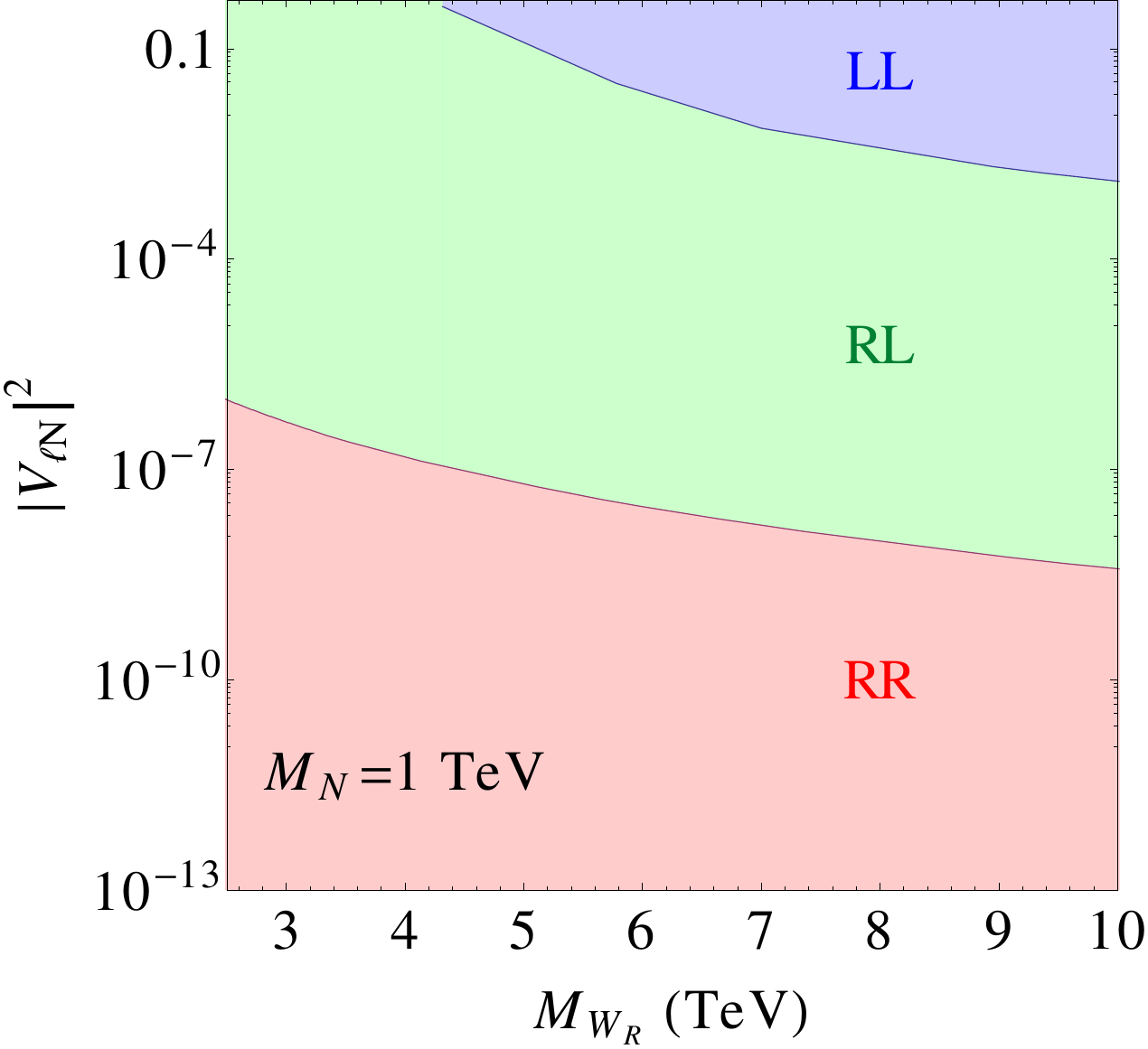}
\hspace{0.0cm}
\includegraphics[width=4.2cm]{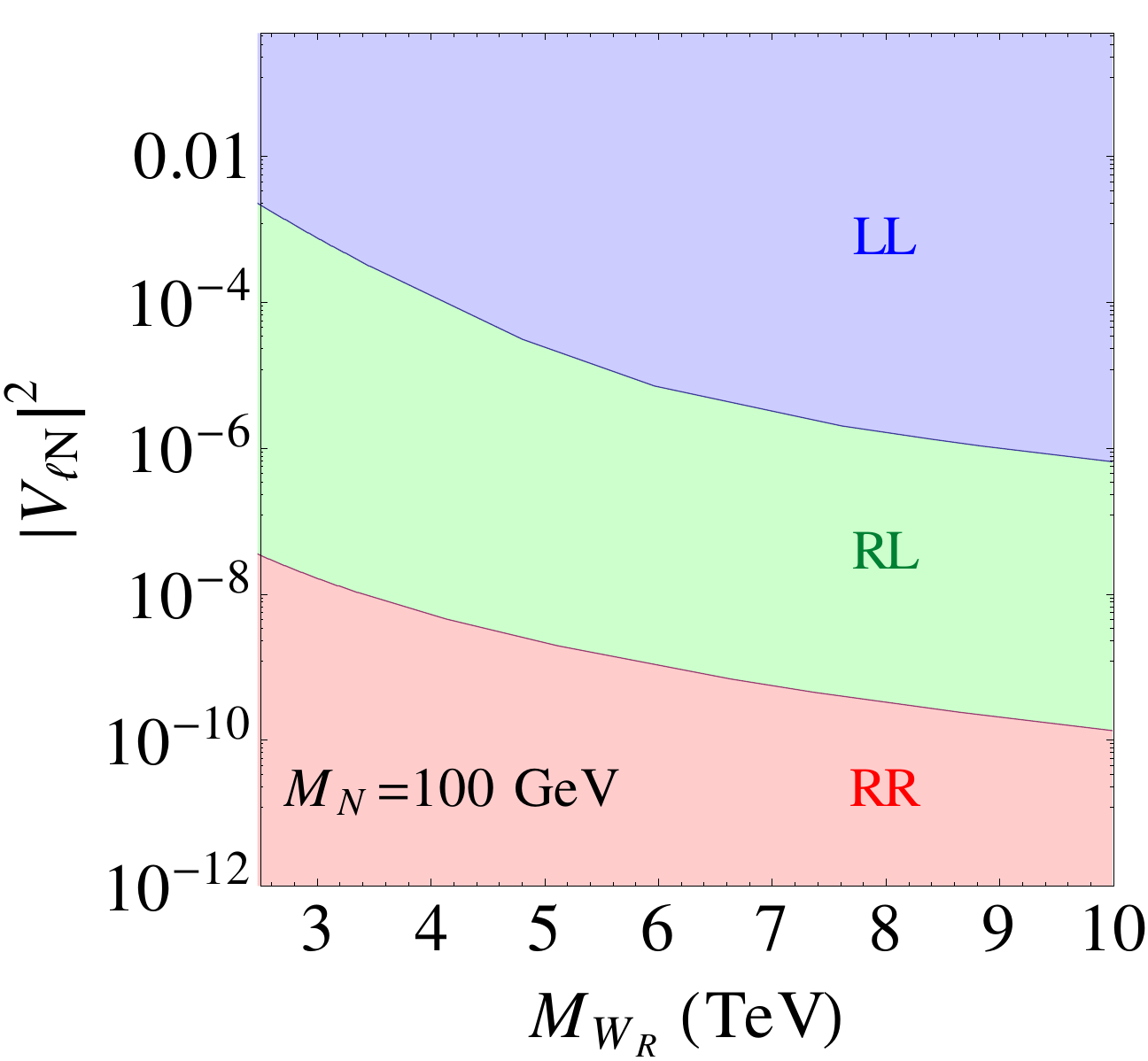}
\caption{The phase diagram for the collider signal of a heavy Majorana neutrino in the minimal L-R seesaw model.}
\label{fig2}
\end{figure}

The regions of dominance for various contributions discussed above are shown in Fig.~\ref{fig2} (the `L-R phase diagram') for two typical choices of the heavy neutrino mass $M_N=100$ GeV and 1 TeV. The different shaded regions show the dominant contributions from RR, RL, and LL channels respectively to the $\ell^\pm\ell^\pm jj$ signal at parton-level. Note that the other remaining possibility, namely, the LR contribution (Fig.~\ref{fig1}d), is doubly suppressed by the mixing as well as phase space, and hence, always smaller than one of the other three contributions shown here.  The RL dominance clearly spans a wide parameter space of the model, and in particular, it can go all the way down to $|V_{\ell N}|\geq 10^{-5}$, close to the `vanilla' seesaw expectation of $\sqrt{M_\nu/M_N}$. 
Thus combining collider studies with fitting the neutrino oscillation data using Eq.~(\ref{seesaw}) which also depends on heavy-light mixing and heavy neutrino mass scale can play a decisive role in testing TeV L-R seesaw models at the Energy frontier.

\subsection{Prospects for improved Collider Limit on the Heavy-Light Neutrino Mixing}
As an immediate consequence of our results discussed above, we can derive 
improved collider limits on the heavy-light neutrino mixing compared to the existing limits~\cite{CMS-LL, ATLAS-LL} which were obtained from $\sqrt s=7$ TeV LHC data 
assuming the inclusive signal cross section for the LL mode alone. 
For the range of mixing parameter being constrained here, the RL contribution 
is in general dominant and the total (LL+RL) inclusive 
cross section is larger thus yielding a stronger limit on the mixing parameter. 
%
To be more precise, for an experimentally observed limit on the signal cross section $\sigma_{\rm expt}$, we can infer the following: (i) the $(M_N,M_{W_R})$ plane for which $\sigma_{\rm RL}\geq \sigma_{\rm expt}$ is ruled out, and (ii) for $\sigma_{\rm RL}<\tilde{\sigma}_{LL}<\sigma_{\rm expt}$ where $\tilde{\sigma}_{LL}\equiv \sigma_{\rm LL}/|V_{\ell N}|^2$ is the normalized LL cross section, the new limit on the mixing parameter will be 
\begin{eqnarray}
|V_{\ell N}|^2 < \frac{\sigma_{\rm expt}-\sigma_{\rm RL}}{\tilde{\sigma}_{\rm LL}}
\label{eq:9}
\end{eqnarray}
 which is obviously stronger than that derived assuming $\sigma_{\rm RL}=0$. For instance, using the observed cross section limit for $\sqrt s=7$ TeV from the ATLAS analysis~\cite{ATLAS-LL}, we obtain using Eq.~(\ref{eq:9}) the upper limit 
on $|V_{\ell N}|^2$ to be  0.0095 (0.1614) at $M_{W_R}=2.5$ TeV and 
$M_N=100~(300)$ GeV, compared to the existing limit of 0.01 (0.18)~\cite{ATLAS-LL}. We expect this improvement to be much more prominent for higher values of 
$M_N$ and/or at $\sqrt s=14$ TeV LHC.
For illustration, assuming the expected upper limit on the signal cross section for $\sqrt s=14$ TeV to be smaller than the observed limit for $\sqrt s=7$ TeV, we obtain the following conservative upper limit on $|V_{\ell N}|^2$ for $M_N=100$ GeV: $1.2\times 10^{-3}$ from the LL channel only, compared to $4.6~(9.3)\times 10^{-4}$ from the (LL+RL) channel for $M_{W_R}=3~(3.5)$ TeV. For comparison, the current best limits on the heavy-light neutrino mixing for the muon and tau sectors come from electroweak precision data: $|V_{\mu N}|^2 < 3.3\times 10^{-3}$ and $|V_{\tau N}|^2 < 6.2\times 10^{-3}$ at 90\% C.L.~\cite{delAguila:2008pw}, whereas for the electron sector, the limit derived from the recent $0\nu\beta\beta$ search results is more stringent: $M_{W_R}^{-4}|\sum_i V^2_{eN_i}/M_{N_i}|<0.1~{\rm TeV}^{-5}$~\cite{dgmr}.   

\begin{figure}[t]
\centering
\includegraphics[width=4cm]{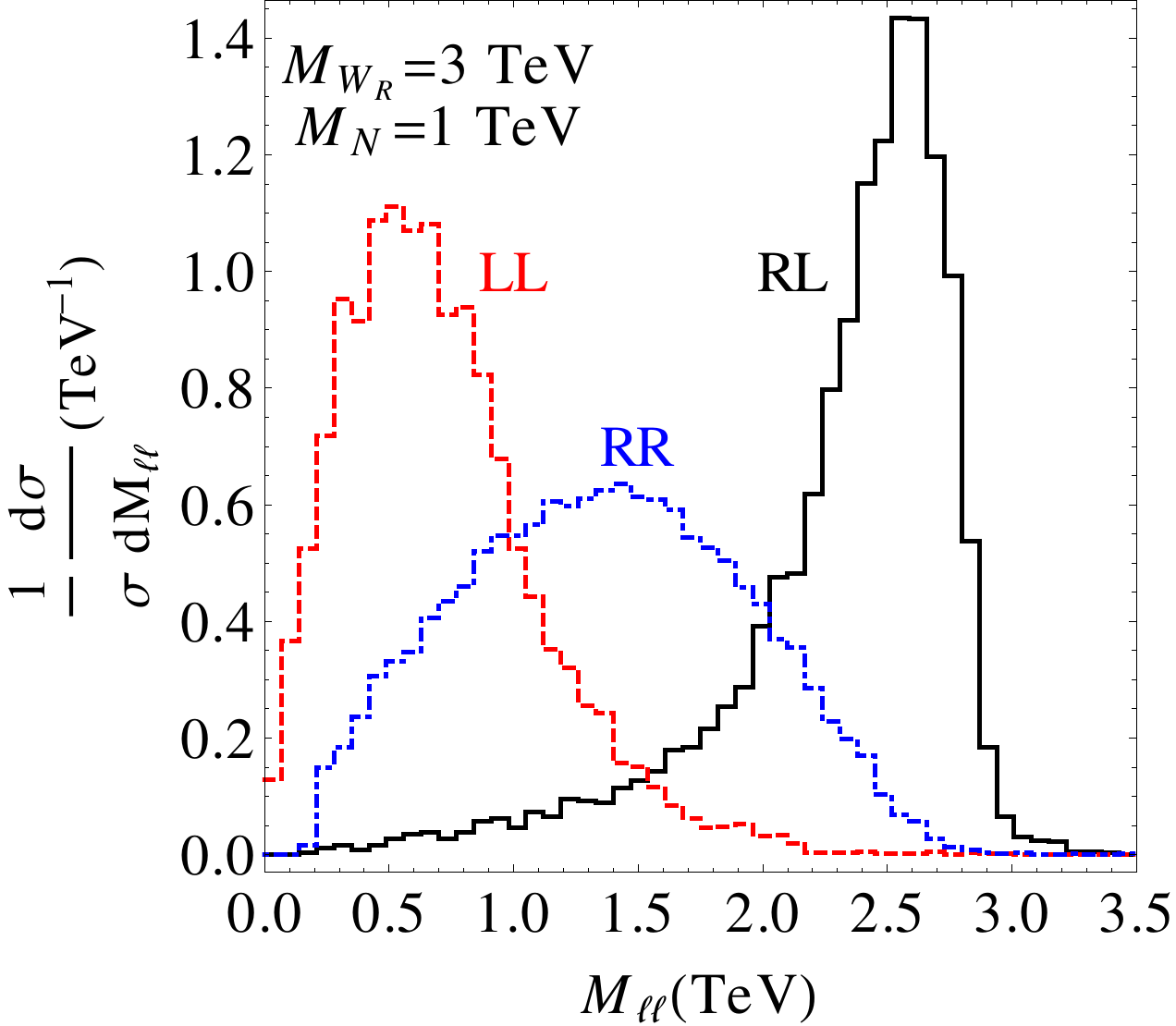}
\hspace{0.0cm}
\includegraphics[width=4cm]{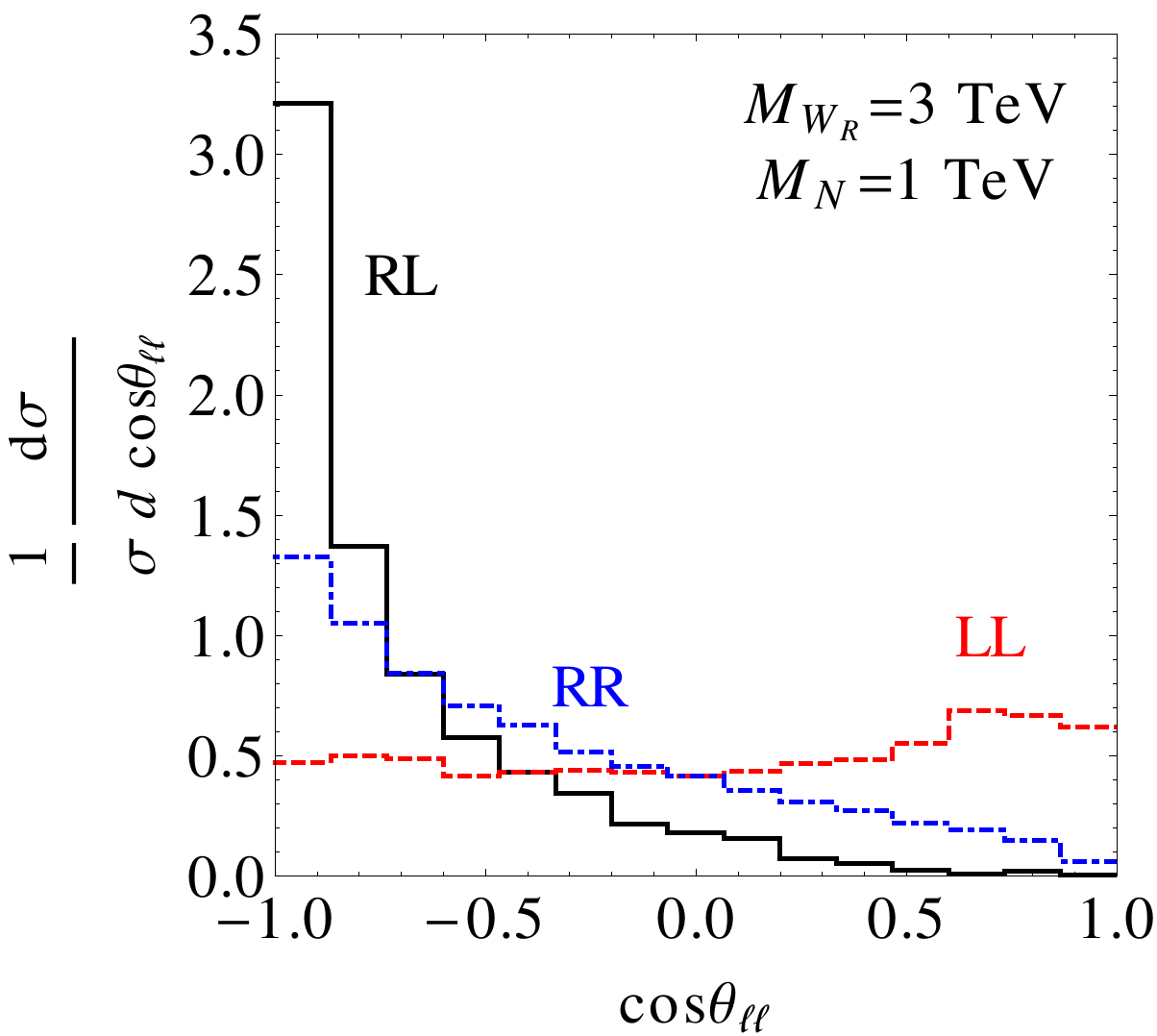}
\caption{The invariant mass distribution and the angular 
correlation of the final-state leptons for the LL, RL and RR modes shown in Fig.~\ref{fig1}. For comparison, all distributions have been normalized to unity.}
\label{fig5}
\end{figure}

Once the $\ell^\pm\ell^\pm jj$ signal is observed 
at the LHC, it is important to distinguish between the various modes shown in Fig.~\ref{fig1} 
 which can be used to prove the existence of a low-scale L-R symmetry.  
Two useful kinematic variables for this purpose are the dilepton invariant mass distribution and the angular correlation between the final-state charged leptons. 
Simulation results discussed in~\cite{chendev} are shown 
in Fig.~\ref{fig5}, after implementing the realistic cuts from existing  
experimental analyses and taking into account the detector effects. 

\section{Lepton Flavor Violation consequences}\label{sec5}
\begin{figure}[t]
\begin{center}
\begin{tabular}{ccc}
\includegraphics[width=3cm]{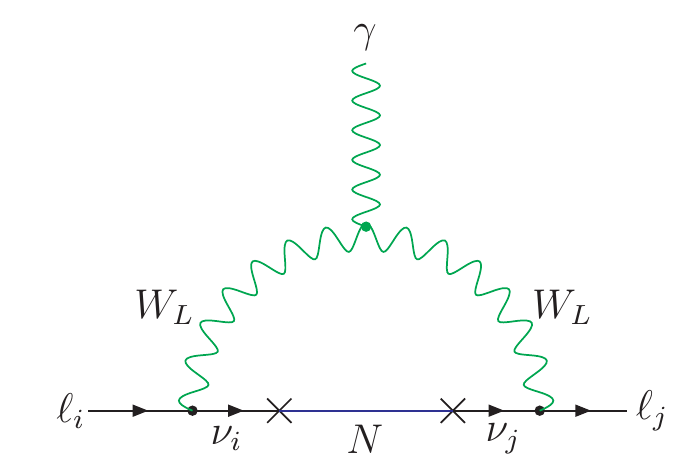} & 
\includegraphics[width=3cm]{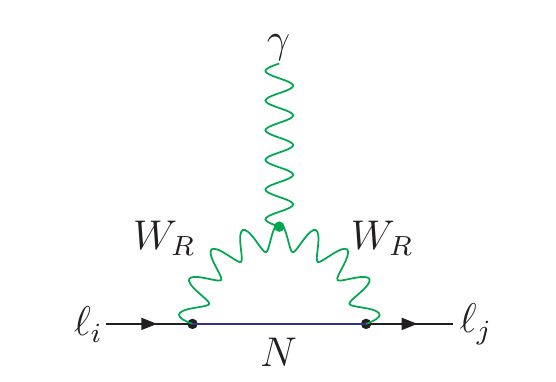} & 
\includegraphics[width=3cm]{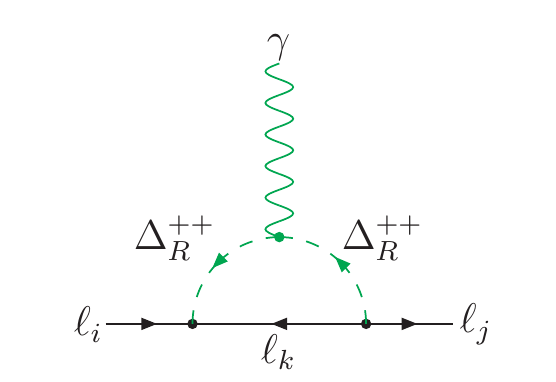}  
\\
(a) & (b)  & (c) 
\end{tabular}
\end{center}
\caption{Various one-loop diagrams contributing to the LFV process $\ell_i\to \ell_j\gamma$ in the minimal L-R seesaw model.}\label{fig6}
\end{figure}

In the `vanilla' as well as the special class of TeV scale L-R seesaw models discussed above, the LFV processes receive new contributions from the RH currents.
In the minimal SM-seesaw case, the only class of graphs that lead to enhanced LFV signal (e.g., in $\mu\to e\gamma$ process) arises from heavy-light mixing in second order and involves the $W_L$ exchange~\cite{mueg-L} (see Fig.~\ref{fig6}a). In cases where $m_D$ has special forms that lead to large $V_{\ell N}$, it has been noted that BR($(\mu\to e\gamma)$ can be as large as $10^{-13}$ for $M_N\leq 200$ GeV, whereas BR$(\tau\to\mu\gamma)$ can be as large $10^{-9}$~\cite{Alonso:2012ji, dlm}. All these are in the observable range of current and planned experiments~\cite{meg, future}. 

In the L-R model, new contributions to $\mu\to e\gamma$ can arise from different sources depending  on the details of the model. In the `vanilla' model (model A), the $W_R$ - $N$ virtual state (see Fig.~\ref{fig6}b) gives a new contribution which scales like $M^{-8}_{W_R}$ in the branching ratio~\cite{riaz}:
\begin{eqnarray}
{\rm BR}({\mu\to e\gamma})_{W_R}\simeq \frac{3\alpha}{32\pi}\left(\frac{M_{W_L}}{M_{W_R}}\right)^8 \left(s_R c_R\frac{M^2_{N_2}-M^2_{N_1}}{M^2_{W_L}}\right)^2 \nonumber \\
\end{eqnarray}
where $s_R, c_R$ are the mixings in the RH charged leptonic current interaction. The interesting aspect of this diagram is that it only depends on the mixings in the  RH charged current interaction with $W_R$  in a manner analogous to the well known  GIM mechanism~\cite{gim} in the SM. In particular, it is independent of observed neutrino mixings. For a hierarchical RH neutrino spectrum with $M_{N_2}=1~{\rm TeV}\gg M_{N_1}$ and for maximal mixing angle $\theta_R=\pi/4$, the current upper limit on BR$(\mu\to e\gamma)<5.7\times 10^{-13}$ at 90\% CL~\cite{meg} implies that $M_{W_R}>3.4$ TeV. When BR$({\mu\to e\gamma})$ is searched down to $10^{-16}$ level~\cite{future}, it can probe $M_{W_R}$ up to 10 TeV. Thus if there are no flavor symmetries endowing special structure to the RH neutrino mass matrix, this can be a potent way to shed light on this class of TeV seesaw models.

Other contributions to $\mu\to e\gamma$  process come from the scalar sector of the L-R model involving $\Delta^{++}_R$ fields in the loop~\cite{rnm-92}:
\begin{eqnarray}
{\rm BR}({\mu\to e\gamma})_{\Delta_R^{++}}\simeq \frac{2\alpha M^4_{W_L}}{3\pi g^4}\left[\frac{(f^\dagger f)_{12}}{M^2_{\Delta^{++}}}\right]^2
\end{eqnarray} 
This has been calculated for model A under certain assumptions in~\cite{ciri, tello} and the current MEG limit~\cite{meg} implies a lower  bound on $M_{\Delta^{++}}\geq 1.7$ TeV for RH charged current mixing $\sim 0.01$. 
In model B discussed in Section~\ref{sec2}, where all RH neutrino masses and mixing angles are fixed by the neutrino mass fit, we estimate this contribution to BR(${\mu\to e\gamma})$ to be $\sim 3\times 10^{-15}$ for $M_{\Delta^{++}}\simeq 1 $ TeV and $M_{W_R}\simeq 5$ TeV~\cite{dlm}.

Other LFV processes such as $\mu$ to $e$ conversion and $\mu\to 3e$ also receive several new contributions in the TeV L-R model. For instance, the $\mu\to 3e$ process gets contributions from a photon mediated one loop graph for $\mu\to e\gamma$ in Fig.~\ref{fig6} with a virtual $\gamma$, loop box graphs with $W_R$ and $N$ virtual states plus a tree-level graph involving the exchange of $\Delta^{++}_{R,L}$ states. The generic formula for the tree-level $\Delta$ graph is given by~\cite{pal}
\begin{eqnarray}
{\rm BR}({\mu\to 3e})\simeq \frac{1}{2}\left(\frac{M_{W_L}}{M_{W_R}}\right)^4\left(\frac{M_{N,12}M_{N,11}}{M^2_{\Delta^{++}_R}}\right)^2
\end{eqnarray}
In model B, since the neutrino mass fit fixes all the parameters of the model except $M_{W_R}$ and $M_{\Delta^{++}}$, for $M_{W_R}=3$ TeV and $M_{\Delta^{++}_R}=1$ TeV, we predict BR$(\mu\to 3e)\simeq 3\times 10^{-13}$ which is only a factor of 3 smaller than the current upper bound~\cite{PDG}. This could be used as a test of the model B in~\cite{dlm}.

\section{Implications for neutrinoless double beta decay}\label{sec6}
In this section, we discuss tests of the TeV L-R seesaw model in $0\nu\beta\beta$ process. Since in L-R seesaw models both the light ($\nu_e$) and heavy neutrinos ($N$) are Majorana fermions, they break lepton number by two units and lead to the classic $0\nu\beta\beta$ process $(A,Z)\to (A, Z+2)+e^-+e^-$~\cite{rode}. The first contribution to this process comes from the well known light neutrino exchange and the amplitude $A_{{0\nu}\beta\beta}$ is proportional to $G^2_Fm_{\nu_e}$, where $G_F$ is the Fermi coupling constant. The heavy neutrino contribution to $A_{{0\nu}\beta\beta}$ is given by $\sim G^2_F\left(M_{W_L}/M_{W_R}\right)^4/M_N$. Current lower limits~\cite{0nbbex} on the half-life of this process already constrain the parameter space of the minimal TeV L-R model:
\begin{eqnarray} 
M^{1/4}_N M_{W_R} \geq (11~{\rm TeV})^{5/4}
\end{eqnarray}
There are also other contributions of similar order of magnitude coming from $\Delta^{++}_R$ exchange~\cite{MV}. Using the recent experimental limits, one can derive the following lower bound on the mass of $\Delta^{++}_R$:
\begin{eqnarray}
M_{\Delta^{++}_R}\geq (500~{\rm GeV}) \left(\frac{3.5~{\rm TeV}}{M_{W_R}}\right)^2\left( \frac{M_N}{3~{\rm TeV}}\right)^{1/2}
\end{eqnarray}
Note that the LFV constraints seem to imply $M_{N}/M_\Delta \leq 0.1$~\cite{tello}. Similarly for large heavy-light mixing, the so-called $\lambda$- and $\eta$-diagrams~\cite{hirsch} could contribute significantly to $0\nu\beta\beta$~\cite{barry}, and must be taken into account in a complete analysis. 
As the search for $0\nu\beta\beta$ becomes more and more sensitive in near future, it will probe a wider range parameters of the TeV L-R seesaw model. 
It turns out that due to the particular structure of the RH neutrino mass matrix in model B, the $0\nu\beta\beta$ process is suppressed~\cite{dlm} and is about an order of magnitude (in amplitude) beyond the reach of current experiments.

\section{Conclusion}\label{conc}
We have given a brief overview of the possible ways to further explore the TeV Left-Right seesaw model and thereby shed light on the origin of neutrino masses both at the Energy frontier (LHC) and at the Intensity frontier.  We have emphasized  a recently noted  new contribution to the smoking gun collider signals of TeV-scale L-R seesaw coming from the heavy-light neutrino mixing contribution  (called RL in the text) which can dominate over the usually discussed  right-handed and left-handed charged current contributions.  Probing this new contribution at the LHC can provide crucial information about the detailed nature of the seesaw mechanism.  
We also discuss a class of models where such enhanced heavy-light neutrino mixings can arise. The generic and special TeV scale L-R seesaw models also have a plethora of new tests 
at the Intensity frontier, as higher intensity of lepton sources come online. These tests including searches for various charged  lepton flavor violating processes such as $\mu\to e\gamma$ and $\mu\to 3e$ as well as for the lepton number violating process $0\nu\beta\beta$ will improve our knowledge about the L-R symmetric realization of seesaw at the TeV scale. 

\section*{Acknowledgments}
RNM would like to acknowledge useful discussions with Goran  Senjanovi\'c and Vladimir Tello. We thank Chang-Hun Lee for carefully reading the manuscript. 
The work of PSBD is supported by the 
Lancaster-Manchester-Sheffield Consortium for Fundamental Physics under STFC grant ST/J000418/1, and RNM is supported by National Science Foundation grant No.
PHY-0968854.

\end{document}